\documentclass[letter]{aa}  

\usepackage{graphicx}
\usepackage[normalem]{ulem}
\usepackage{txfonts}
\usepackage{array}
\usepackage{float}
\newcommand{\kms}{$\mathrm{km} \, \mathrm{s}^{-1}$}
\newcommand{\jykms}{$\mathrm{Jy}\:\mathrm{km}\,\mathrm{s}^{-1}$}

\newcommand{\fgas}{$M_\mathrm{HI}/M_\ast$}
\newcommand{\dg}{DGSAT I}
\newcommand{\mone}{M-161-1}
\newcommand{\rone}{R-127-1}
\newcommand{\sdone}{SdI-1}
\newcommand{\sdtwo}{SdI-2}
\newcommand{\utwentyone}{UGC 2162}

\begin{document}

\title{The HI content of isolated ultra-diffuse galaxies: A sign of multiple formation mechanisms?}

\author{
E. Papastergis\inst{\ref{kapteyn}}\fnmsep\thanks{\textit{NOVA} postdoctoral fellow},
E.A.K. Adams\inst{\ref{astron}}
\and A.J. Romanowsky\inst{\ref{sjsu}}
}

\institute{
Kapteyn Astronomical Institute, University of Groningen, Landleven 12, Groningen NL-9747AD, The Netherlands \\ \email{papastergis@astro.rug.nl}\label{kapteyn}
\and
ASTRON, the Netherlands Institute for Radio Astronomy, Postbus 2, Dwingeloo NL-7900AA, The Netherlands \\ \email{adams@astron.nl}\label{astron}
\and
Department of Physics \& Astronomy, San Jos\'{e} State University, One Washington Square, San Jose, CA 95192, USA \\ \email{aaron.romanowsky@sjsu.edu} \label{sjsu}
%\and
%University of California Observatories, 1156 High Street, Santa Cruz, CA 95064, USA \label{uco}
}

\titlerunning{HI content of isolated UDGs}

\authorrunning{Papastergis, Adams and Romanowsky}

\abstract{
We report on the results of radio observations in the 21 cm emission line of atomic hydrogen (HI) of four relatively isolated ultra-diffuse galaxies (UDGs): \dg, \rone, \mone, and SECCO-dI-2. Our Effelsberg observations resulted in non-detections for the first three UDGs, and a clear detection for the last. \dg, \rone, and \mone \ are quiescent galaxies with gas fractions that are much lower than those of typical field galaxies of the same stellar mass. On the other hand, SECCO-dI-2 is a star forming gas-rich dwarf, similar to two other field UDGs that have literature HI data: SECCO-dI-1 and \utwentyone. This group of three gas-rich UDGs have stellar and gaseous properties that are compatible with a recently proposed theoretical mechanism for the formation of UDGs, based on feedback-driven outflows. In contrast, the physical characteristics of \rone \ and \mone \ are puzzling, given their isolated nature. We interpret this dichotomy in the gaseous properties of field UDGs as a sign of the existence of multiple mechanisms for their formation, with the formation of the quiescent gas-poor UDGs remaining a mystery.
}

   %\keywords{ --  -- }

   \maketitle
%
%________________________________________________________________

\section{Introduction}
\label{sec:intro}

The past two years have witnessed a surge of interest in the study of galaxies characterized by extremely low surface brightness (LSB). Even though LSB galaxies have been studied for decades \citep[][to name just a few]{Impey1988,Dalcanton1997,deBlokMcGaugh1997}, the recent discovery of ultra-diffuse galaxies (UDGs) in the Coma cluster by \citet{vDokkum2015a,vDokkum2015b} has drawn much attention from both observers and theorists. Ultra-diffuse galaxies are broadly defined as galaxies with optical luminosities typical of dwarf galaxies ($L \sim 10^7 - 10^8 \; L_\odot$), but half light radii typical of much larger spirals, such as the Milky Way ($r_e \sim 1.5 - 5$ kpc). After their initial discovery in Coma \citep{vDokkum2015a,Koda2015} UDGs were not only detected in other nearby clusters \citep{Mihos2015,Munoz2015}, but also in lower density environments such as the outskirts of clusters \citep{Martinez2016}, galaxy groups \citep{Makarov2015,RomanTrujillo2016,Trujillo2017}, and even in the field \citep{Bellazzini2017}.

Owing to the extremely low surface brightness ($\mu_{eff,V} \gtrsim 24.5$ mag arcsec$^{-2}$) and red optical colors, cluster UDGs have been conjectured to be ``failed'' galaxies, since their overall stellar content is much lower than that of normal quiescent galaxies of the same linear size. In fact, recent estimates of the dynamical mass of UDGs indicate dynamical-to-stellar mass ratios that are much higher than expected based on the luminosity of the galaxies. One remarkable case is Dragonfly-44, a Coma UDG with a luminosity of $L_V = 2 \times 10^8 L_\odot$. The measured velocity dispersion and globular cluster count of this UDG suggest a host halo mass of $M_h \sim 10^{12} \; M_\odot$, i.e., similar to the halo of the Milky Way (\citealp{vDokkum2016}; but see also \citealp{diCintio2017}). \citet{Beasley2016} and \citet{BeasleyTrujillo2016} have also inferred very high dynamical-to-stellar mass ratios for the UDGs VCC 1287 and Dragonfly-17, but argue that their host halos have dwarf-scale masses ($M_h \sim 10^{11} \; M_\odot$). Further evidence for dwarf-scale host halos has been obtained from stacked globular cluster counts of 18 Coma UDGs \citep[][]{Amorisco2016}. 

Two theoretical formation mechanisms for UDGs have recently been proposed in the literature, whereby UDGs correspond to dwarf-scale halos with unusually extended stellar disks. \citet{AmoriscoLoeb2016} propose that UDGs are simply dwarf galaxies hosted by halos belonging to the high-end tail of the spin distribution (see also \citealp{Rong2017}). In this scenario, field UDGs are expected to be gas rich, since high halo spin seems to facilitate the formation of galaxies with large gas reservoirs \citep[e.g.,][]{Huang2012,Papastergis2013,KimLee2013,Hallenbeck2014,Maddox2015}. Alternatively, \citet{diCintio2017} argue that star formation feedback in some dwarf halos can create an extended, low surface brightness stellar disk, in conjunction with core creation in their dark matter mass profiles \citep[e.g.,][]{Governato2010}. The \citet{diCintio2017} model makes concrete quantitative predictions for the atomic gas content of isolated UDGs, based on the analysis of the NIHAO hydrodynamical simulation suite \citep{Wang2015}. Isolated UDGs should have atomic hydrogen (HI) masses in the range $M_\mathrm{HI} \sim 10^7 - 10^9 \; M_\odot$, with a characteristic correlation whereby more extended UDGs have higher gas fractions and younger stellar populations. Crucially, UDGs are first formed in moderate density environments as gas-rich star-forming dwarfs, according to both formation models, and can later undergo a process of gas removal and star formation quenching if accreted onto denser structures.

As a result, determining the HI masses of isolated UDGs is crucial for understanding their true nature, and for testing proposed formation scenarios. Our knowledge of the HI content of isolated UDGs is still very limited. Only a handful of UDGs located in relatively low-density environments have prior measurements of their HI mass. \dg \ is a red quiescent UDG in the Pisces-Perseus filament \citep{Martinez2016}, which has an upper limit on its atomic hydrogen mass of $M_\mathrm{HI} < 6.3 \times 10^{8} \; M_\odot$ \citep{GiovanelliHaynes1989}. \utwentyone \ \citep{Trujillo2017} is a nearby blue and star-forming UDG in the M77 group with a HIPASS detection yielding $M_\mathrm{HI} = 1.9 \times 10^8 \; M_\odot$ \citep{Meyer2004}. SECCO-dI-1 (hereafter \sdone; \citealp{Bellazzini2017}) is an isolated star-forming UDG with an HI mass of $M_\mathrm{HI} = 1.2 \times 10^9 \; M_\odot$ \citep{Roberts2004} and an extremely high gas fraction, \fgas$\approx 100$. Since this article has been submitted, our knowledge of gas-bearing isolated UDGs has been significantly broadened by the publication of 115 UDGs detected by the ALFALFA blind HI survey \citep{Leisman2017}.

In this letter, we present new radio observations in the 21 cm emission line of HI of four isolated UDGs, \dg, \mone, \rone \ \citep{Dalcanton1997}, and SECCO-dI-2 (hereafter \sdtwo; \citealp{Bellazzini2017}), taken with the Effelsberg radio telescope. The first three objects represent all known quiescent UDGs that are relatively isolated, while the last object is an optically identified star-forming UDG with no prior information on its HI content. The present data, together with some literature results, represent a first attempt to gain a comprehensive view of the atomic gas content of isolated UDGs. The article is organized as follows: In Section \ref{sec:observations} we briefly describe the Effelsberg observations. In Section \ref{sec:results} we present our HI spectra and our results regarding the HI content of the four isolated UDGs. We conclude in Section \ref{sec:discussion} by discussing the significance of our results for proposed mechanisms of UDG formation.

\section{Radio observations of isolated UDGs}
\label{sec:observations}

We have observed \dg, \mone, \rone, and \sdtwo \ in the HI line with the Effelsberg radio telescope, as part of project 111-16. Observations took place on 2-3 February 2017, using the central pixel only of the 7-pixel receiver in the L band. We used a 100 MHz bandwidth divided into $65\,536$ channels, resulting in a native spectral resolution of 1.53 kHz ($\approx\,$0.3 \kms). We observed in position switching mode with on-off subscans of 90 seconds each. We examined the data and dropped subscans with poor data quality or strong radio frequency interference (RFI) near the expected recessional velocity of our sources.

The four galaxies were observed for varying amounts of time to obtain comparable limits in the gas fraction, \fgas, in case of non-detections. The final on-source integration times were 345 mins for \mone, 180 mins for \rone, and 142.5 mins for \dg. The integration time for \sdtwo was much shorter because this source was detected at high signal to noise after only 45 mins of on-source observing time.

\section{Results}
\label{sec:results}

\begin{figure}[!ht]
\centering
\includegraphics[width=\columnwidth,clip=True,trim=0cm 0.5cm 0cm 2cm]{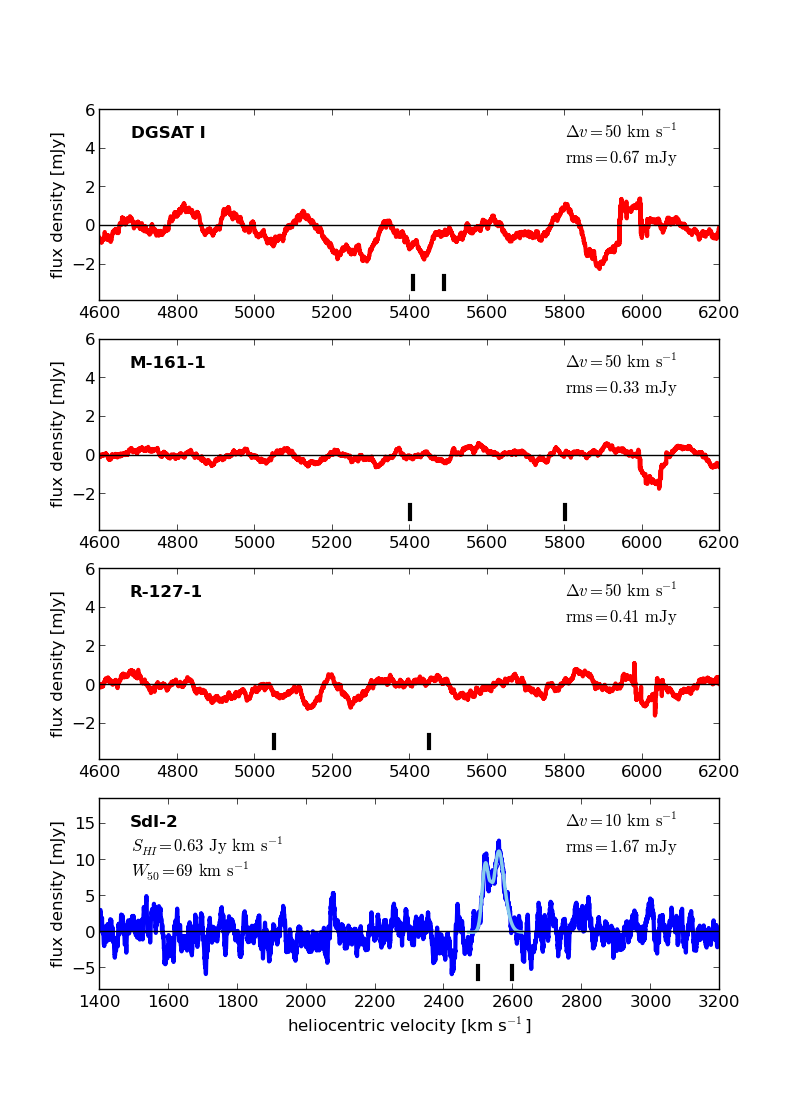}
\caption{
%%title
HI spectra of four isolated UDGs, obtained with the Effelsberg radio telescope. 
%%body:
From top to bottom, the panels correspond to \dg, \mone, \rone, and \sdtwo. The first three spectra are non-detections. \sdtwo \ is instead clearly detected in the last spectrum. The light blue solid line is the best fitting generalized busy function profile \citep{Westmeier2014}. 
In all spectra, the short vertical lines denote the expected systemic velocity range from prior optical redshift measurements.
%Note that the spectra of non-detections are smoothed to a resolution of $\Delta v = 50$ \kms, while the spectrum of SdI-2 is smoothed to $\Delta v = 10$ \kms. Note also that the $x$-axis and $y$-axis range is the same for the three non-detections, but different for SdI-2.  
}
\label{fig:spectra}
\end{figure}

Our Effelsberg observations of \dg, \rone, and \mone \ resulted in HI non-detections, as shown in the top three panels of Figure \ref{fig:spectra}. This outcome is consistent with the fact that these three UDGs all have passive optical spectra. Our spectra can nonetheless be used to refine the existing upper limit on the HI mass of \dg, and to derive the first upper limits on the HI masses of \rone \ and \mone. In spectral line observations, the derived upper limit value depends on the assumed velocity width of the HI profile of the source. In general, a smaller profile width leads to a more stringent upper limit. We adopt here a fiducial value of $W_{50} = 50$ \kms, which approximately corresponds to the typical velocity width of dwarfs with $L_V \sim 10^8 \; L_\odot$ \citep{Ponomareva2017}.We then follow a matched filtering approach, whereby we smooth the spectrum to a velocity resolution that equals the assumed galactic profile width.

We measure the rms fluctuations of the three smoothed spectra to derive $5\sigma$ upper limits to the HI flux as

\begin{eqnarray}
S_\mathrm{HI,lim}\,(\mathrm{mJy}\:\mathrm{km}\,\mathrm{s}^{-1}) \; = 5 \: \times \: rms_{50}\,(\mathrm{mJy}) \: \times 50 \; \mathrm{km}\,\mathrm{s}^{-1} \;\; . 
\end{eqnarray}

\noindent
We then convert these flux upper limits into upper limits in HI mass, under the standard assumption of optically thin emission,

\begin{eqnarray}
M_\mathrm{HI,lim}(M_\odot) = 235.6  \: \times \: S_\mathrm{HI,lim}\,(\mathrm{mJy}\:\mathrm{km}\,\mathrm{s}^{-1}) \: \times \: D\,(\mathrm{Mpc})^2 \;\; .
\end{eqnarray}

\noindent
In the equation above, $D$ is the distance to the source; here we adopt the values 78 Mpc, 75 Mpc, and 81 Mpc for \dg, \rone, and \mone, respectively \citep{Martinez2016,Dalcanton1997}. The 5$\sigma$ upper limits derived from our observations are summarized in Table \ref{tab:udg_data}, and correspond to $M_\mathrm{HI} < \; 2.4, 1.3, 1.3 \; \times 10^8 \; M_\odot$ respectively for \dg, \rone, and \mone. By combining these HI mass limits with estimates of their stellar masses listed in Table \ref{tab:udg_data}, we derive upper limits on their gas fractions of \fgas$< \; 0.61,0.41,0.52$.

\begin{table*}[htb]
    \centering
    \small
    \begin{tabular}{lcccccccccc}
        
        \hline
        
         & (1) & (2) & (3) & (4) & (5) & (6) & (7) & (8) & (9) & (10)  \\
        
        name & type & $D$ & $r_\mathrm{eff}$ & $S_\mathrm{HI}$ & $M_\mathrm{HI}$ & $V_\mathrm{sys}$ & $W_{50}$ & $M_\ast$ & \fgas \ & reference  \\
        
 & & (Mpc) & (kpc) & (\jykms) & ($M_\odot$) & (\kms) & (\kms) & ($M_\odot$) &     \\ [0.5ex]

        \hline \hline 
        
        \multicolumn{9}{l}{\textbf{this work}}\\ [0.5ex]
        
        \hline

        \dg \ & q & 78 & 4.5 & $<0.17$ ($5\sigma$) & $< 2.4 \times 10^8$ ($5\sigma$) & 5450 $\pm$ 40$^a$ & ... & $4 \times 10^8$ & $<0.61$ ($5\sigma$) &  M16 \\
        
        \mone \ & q   & 81 & 4.1 & $<0.083$ ($5\sigma$) & $< 1.3 \times 10^8$ ($5\sigma$) & 5600 $\pm$ 200$^a$ & ... & $2.5 \times 10^8 \: ^b$ &  $< 0.52$ ($5\sigma$) & D97   \\

        \rone \ & q   & 75 & 4.2 & $<0.10$ ($5\sigma$) & $< 1.3 \times 10^8$ ($5\sigma$) & 5250 $\pm$ 200$^a$ & ... & $3.2 \times 10^8\: ^b$ &  $<0.41$ ($5\sigma$) & D97   \\

        \sdtwo \ & sf   & 40 & 1.3 & 0.63 & $2.4 \times 10^8$ & 2543 & 69 & $0.9 \times 10^7$ &  27  & B17   \\

        \hline 
        
        \multicolumn{9}{l}{\textbf{from literature}} \\
        [0.5ex]
        
        \hline 
        
        \sdone \ & sf  & 112 & 2.6 &  & $1.2 \times 10^9$ & 7791  & 90 & $1 \times 10^7$ &  120 & B17,R04    \\
        %$\pm$ 13

          \utwentyone \ & sf  & 12.3 & 1.7 &  & $1.9 \times 10^8$ & 1172 & 55 & $2 \times 10^7$ & 10 & T17,M04     \\

        %40 from ALFALFA

        \hline

    \end{tabular}
    
    \caption{Gaseous and stellar properties of six isolated UDGs. (1) Quiescent (q) or star-forming (sf) galaxy according to optical spectrum. (2) Distance. (3) Radius enclosing half of the total light. (4) Total flux of the HI emission line. Reported only for objects observed in this work. (5) Total HI mass. (6) Heliocentric recessional velocity, measured from the central velocity of the HI line profile. (7) Observed velocity width of the HI line profile, at 50\% of the peak intensity level (uncorrected for inclination). (8) Stellar mass. (9) HI gas fraction. (10) References: \citet[M16]{Martinez2016}, \citet[D97]{Dalcanton1997}, \citet[B17]{Bellazzini2017}, \citet[T17]{Trujillo2017}, \citet[R04]{Roberts2004}, \citet[M04]{Meyer2004}.\hspace*{0.75cm}$^a$Redshifts from optical spectra. $^b$Stellar masses are calculated from the $V$-band magnitude (D97), and assuming $V-I = 1.0$ (same color as \dg). We use the mass-to-light calibration of \citet[Table 3]{IntoPortinari2013}.}
    \label{tab:udg_data}

\end{table*}

In contrast, the Effelsberg spectrum of \sdtwo \ reveals a clear detection (bottom panel of Fig. \ref{fig:spectra}). We fit a generalized busy function to the HI profile of \sdtwo \ \citep[\S4.1]{Westmeier2014}, which results in a flux of $S_\mathrm{HI} = 0.63$ \jykms \ and corresponding HI mass at a distance of 40 Mpc of $M_\mathrm{HI} = 2.4 \times 10^8 \; M_\odot$ . Moreover, the HI profile of \sdtwo \ has the characteristic double-horned shape with a velocity width projected on the line of sight of $W_{50} = 69$ \kms. Owing to the low stellar mass of \sdtwo \ ($M_\ast \approx 10^7 \; M_\odot$), this UDG has a high gas fraction of $M_\mathrm{HI}/M_\ast = 27$. As a result, \sdtwo \ is very similar to two other relatively isolated UDGs, \sdone \ and \utwentyone, with HI measurements in the literature ($M_\ast \approx 10^7 \; M_\odot$, \fgas$ \gg 1$; see Table \ref{tab:udg_data}). These three gas-rich UDGs are also similar to each other in terms of optical properties, as they all have emission line spectra \citep{Bellazzini2017,Trujillo2017}.

\section{Discussion}
\label{sec:discussion}

\begin{figure}[!ht]
\centering
\includegraphics[width=\columnwidth,clip=True,trim=0cm 0cm 0cm 1.5cm]{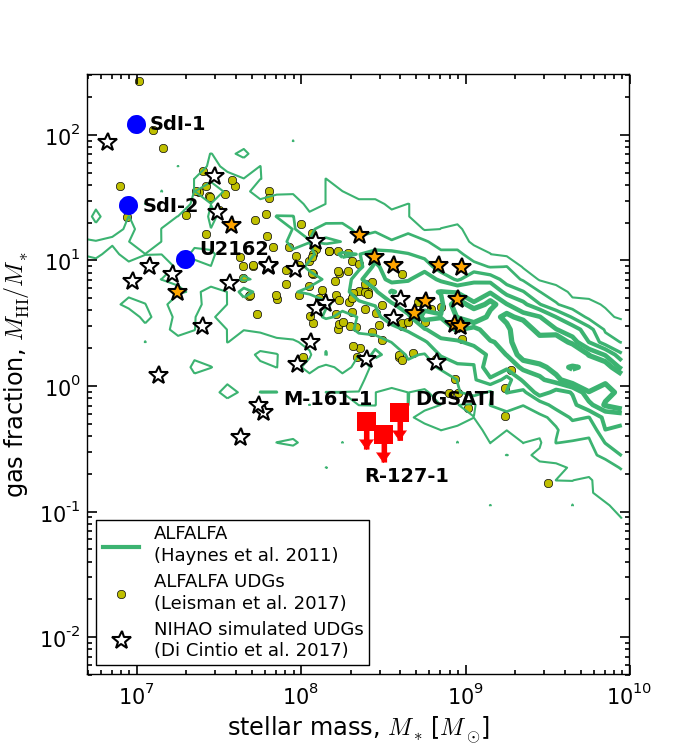}
\caption{
%%title
Position of isolated UDGs on the stellar mass--gas fraction plane. 
%%body: 
The large symbols correspond to six known isolated UDGs, four of which have been observed in HI as part of this work (see Table \ref{tab:udg_data}). The red squares correspond to HI upper limits, while the blue circles correspond to HI detections. The solid contours represent galaxies detected by the ALFALFA blind HI survey \citep[$\alpha$.40 catalog;][]{Haynes2011}. 
The lowest contour encloses 95\% of the ALFALFA detections and each successive contour encloses 15\% less. 
Gas-bearing UDGs detected by the ALFALFA survey are denoted with small yellow circles \citep[][]{Leisman2017}.
Star symbols are simulated field UDGs from the NIHAO simulation sample \citep{Wang2015}, as presented in \citet{diCintio2017}. Those with an orange filling represent the most extended objects ($r_\mathrm{eff} > 3$ kpc).
%The dotted ellipse represents the 1$\sigma$ error assuming 0.3 dex uncertainty in the stellar mass-to-light ratio, 5\% uncertainty in the distance, and 10\% uncertainty in the HI flux calibration.      
%%conclusion:
Please refer to Sec. \ref{sec:discussion} for the scientific interpretation of this figure.
%Three of these six isolated UDGs are objects with gas fractions typical of late-type dwarfs, while the other three are relatively gas-poor objects that are unusual in the field environment. 
}
\label{fig:mstar_vs_fgas}
\end{figure}

The gas fraction upper limits derived for \dg, \rone, and \mone \ do not prove that these three UDGs are truly gas-poor objects. For example, the average gas fraction of dwarf ellipticals in the Virgo cluster is \fgas$<0.025$ \citep{Hallenbeck2012}. As evident in Figure \ref{fig:mstar_vs_fgas}, however, they are sufficiently stringent to demonstrate that these UDGs have less atomic gas than the overwhelming majority of field dwarfs with similar stellar masses detected by the ALFALFA blind HI survey \citep{Haynes2011,Huang2012}. We verified that this is also the case when the \citet{Du2015} subsample of LSB galaxies within ALFALFA is considered\footnotemark{}.

\footnotetext{Stellar masses for ALFALFA galaxies were derived from SED-fitting of pipeline SDSS photometry \citep{Huang2012}, while for the ALFALFA LSB galaxies stellar masses are calculated from reprocessed SDSS photometry in the $g$ and $r$ bands \citep[][]{Du2015} and the mass-to-light calibration of \citet[Table A1]{RoedigerCourteau2015}. }

 The result above is puzzling given the environment in which these UDGs are situated. This is especially true in the case of \rone \ and \mone, which are typical field dwarfs with no massive neighbors ($M_\ast \gtrsim 10^{10} \; M_\odot$) within 1.5 Mpc in projected distance. \citet{Geha2012} find that dwarf galaxies in the SDSS spectroscopic sample that are similarly isolated always display signs of star formation in their optical spectra and thus are presumably gas-rich. \rone \ and \mone \ instead have quiescent optical spectra \citep[Fig. 5]{Dalcanton1997} and low gas fractions. Given their isolation, \rone \ and \mone \ should not have experienced strong environmental effects, while at the same time their stellar masses are orders of magnitude above the scale where cosmic reionization feedback can suppress galaxy formation \citep[e.g.,][]{Okamoto2008}.  

In the case of \dg, the local environment may have played a role in shaping the characteristics of the galaxy. More specifically, \dg \ is located in the outskirts of the cluster UGCl 020 (also Zw 0107+3212) in the Pisces-Perseus filament and could potentially be a ``backsplash'' galaxy \citep{Gill2005}. This UDG has an intermediate mass neighbor at a projected distance of 0.6 Mpc (IC 1668 with $M_\ast \approx 4 \times 10^9 \; M_\odot$) and two massive neighbors within 1.0 Mpc (UGC 862 and CGCG 502-039, with $M_\ast \approx 3 \times 10^{10} \; M_\odot$). At the same time, it should be kept in mind that \dg \ is still found in a relatively low density environment, especially when compared to the vast majority of UDGs discovered to date \citep[e.g.,][]{Yagi2016}. In fact, even the star-forming UDG progenitors discovered by \citet{RomanTrujillo2016} in nearby compact groups lie at a projected distance of just 0.2 - 0.3 Mpc from massive neighbors. 

%\footnotetext{In the case of DGSAT I, ``neighbours'' include only objects with recessional velocities within $\pm$500 \kms \ from the optically measured recessional velocity of DGSAT I ($V_\mathrm{DGSATI} = 5450$ \kms; \citealp{Martinez2016}).}

In contrast, Figure \ref{fig:mstar_vs_fgas} shows that \sdone, \sdtwo, and \utwentyone \ have gas fractions that are entirely consistent with an extrapolation of the trend seen for ALFALFA dwarfs at slightly higher stellar masses. In fact, these three gas-rich UDGs seem to be part of the same population of gas-bearing UDGs detected by ALFALFA \citep{Leisman2017}. Figure \ref{fig:mstar_vs_fgas} further shows that the stellar and gaseous masses of gas-bearing UDGs are consistent with the predictions of the theoretical model of UDG formation put forward by \citet{diCintio2017}. According to this model, UDGs correspond to field dwarfs with a particularly extended and bursty star formation history. The star formation bursts lead to repeated episodes of strong galactic outflows, which in turn cause a systematic  expansion of the stellar orbits and the consequent formation of an extended, low surface brightness stellar disk. Given the high gas fractions of \sdone, \sdtwo, and \utwentyone, these objects may also be compatible with formation scenarios involving high spin host halos \citep{AmoriscoLoeb2016,Rong2017}, but it should be kept in mind that concrete predictions for the HI content of field UDGs in these models are not available yet.

Overall, Figure \ref{fig:mstar_vs_fgas} reveals an unexpected dichotomy in the properties of field UDGs. The quiescent UDGs \dg, \rone, and \mone \ are characterized by low gas fractions, which clearly distinguish them from the population of gas-bearing UDGs and normal late-type dwarfs. Figure \ref{fig:mstar_vs_fgas} also shows that the predictions of the \citet{diCintio2017} model do not seem to match the properties of these three quiescent and gas-poor UDGs. More specifically, the model predicts a positive correlation between $r_\mathrm{eff}$ and \fgas. As a result, the simulated UDGs that are as extended as our quiescent UDGs are too gas rich\footnotemark{}. The physical properties of our three quiescent UDGs remain thus difficult to explain. This is especially true for \rone \ and \mone, which are genuine field dwarfs that are \textit{as isolated as} \sdone \ and \sdtwo \ \citep[refer to \S2.1 in][]{Bellazzini2017}. Perhaps subtle environmental effects (e.g., cosmic web stripping; \citealp{Benitez2013}) or alternative internal feedback mechanisms (e.g., early globular cluster formation; \citealp{KatzRicotti2013}) are needed to reproduce the puzzling properties of these galaxies. The puzzle of gas-poor and quiescent field galaxies pertains not only to UDGs, but extends also to fainter dwarfs in the Local Volume \citep[e.g.,][]{Karachentsev2014}. In the future, the present analysis can be improved significantly by assembling a larger sample of optically identified field UDGs with HI follow-up observations, covering a broad range in stellar mass and optical colors.

\footnotetext{Some of the extended and gas-rich simulated UDGs in \citet{diCintio2017} have red optical colors ($B-R > 0.8$). This means that the optical properties of our quiescent UDGs are not sufficient by themselves to make a comparison with the predictions of the model, and thus our gas fraction limits have been necessary for this purpose.}

\begin{acknowledgements}
This article is based on observations with the 100 m telescope of the MPIfR (Max-Planck-Institut f\"{u}r Radioastronomie) at Effelsberg. Our research has received funding from the European
Union's Horizon 2020 research and innovation program under the RadioNet grant agreement (No730562). We would like to personally thank Benjamin Winkel and Alex Kraus for valuable help with observing and data reduction at Effelsberg, as well as Arianna Di Cintio and the NIHAO collaboration for sharing unpublished data on their simulated UDGs. \\
E.P. is supported by a NOVA postdoctoral fellowship at the Kapteyn Astronomical Institute. E.A.K.A. is supported by grant TOP1EW.14.105 of the Netherlands Organisation for Scientific Research (NWO). A.J.R. was supported by US National Science Foundation grant AST-1616710.
\end{acknowledgements}

%-------------------------------------------------------------------

\end{document}